\documentclass[prl,aps,twocolumn,graphicx,nofootinbib]{revtex4}

\usepackage{exscale}
\usepackage{graphicx}
\usepackage{amsmath}
\usepackage{latexsym}
\usepackage{amsfonts}
\usepackage{amssymb}

\begin{document}

\title{Trapped Ion Chain as a Neural Network: Error Resistant Quantum Computation}
\author{Marisa Pons\(^1\), Veronica Ahufinger\(^2\), Christof Wunderlich\(^3\), Anna Sanpera\(^4\), 
Sibylle Braungardt$^*$, Aditi Sen(De)$^*$, Ujjwal Sen$^*$, and Maciej Lewenstein\(^5\)}

\affiliation{\(^1\)Departamento de F\'isica Aplicada I, Universidad del Pa\'is Vasco, 20600 Eibar, Spain.\\
\(^2\)ICREA and Grup d'\`Optica, Departament de F\'isica, Universitat Aut\`onoma de Barcelona, 08193 Bellaterra (Barcelona),  Spain.\\
\(^3\)Fachbereich Physik, Universit\"at Siegen, 57068 Siegen, Germany.\\
\(^4\)ICREA and Grup de F\'isica Te\`orica, Departament de F\'isica, Universitat Aut\`onoma de Barcelona, 08193 Bellaterra  (Barcelona),  Spain.\\
\(^5\)ICREA and $^*$ICFO-Institut de Ci\`encies Fot\`oniques, 
08860 Castelldefels (Barcelona), Spain.}



\begin{abstract}
We demonstrate the possibility of realizing a neural network in a chain
of trapped ions with induced long range interactions. 
Such models permit one to store information distributed over the whole system.
The storage capacity of such network, which depends on the phonon spectrum of the system, can
be controlled by changing the external trapping potential.
We analyze the implementation of 
error resistant universal quantum information processing in such systems.
\end{abstract} 
\maketitle

The Cirac-Zoller proposal of trapped ion computer \cite{cirac} has become one of
the paradigmatic models to implement a quantum
computer \cite{general}. Recently, spectacular experimental progress
in realization of simple algorithms and implementation of
quantum logic has been achieved using a few ionic qubits (c.f.
\cite{experiments}). 
Although the achievement of an all-purpose quantum computer in 
the near future seems 
difficult, one can be quite optimistic about the
applications of chains of trapped ions as quantum simulators.
Recently, it has been shown that long range (LR) pairwise interactions
between individual spins are induced in an ion trap, when applying
an additional state-dependent force acting on the ions
\cite{christoph,Wunderlich02}. Also a state-dependent optical force can
evoke LR couplings and has been 
proposed to simulate spin 1/2 chain systems \cite{porras}. 

Here we show that ion spin systems can serve to realize a neural network (NN) model. NN are a prototype model of
parallel distributed memory \cite{Amit,parisi}, and
have been intensively studied by physicists since the famous paper
by Hopfield \cite{hopfield}. These disordered systems with 
LR interactions,  typically present a large number of metastable
(free) energy minima, like in spin glasses \cite{parisi}.
These states can be used to store information distributed over the
whole system. The patterns stored have large basins
of attraction in the thermodynamical sense,
so that even fuzzy ones are recognized as perfect ones. For this reason,
attractive NN's can be used as associative memory. At the same time,
NN are robust, so that destroying even a large part of the network
does not necessarily diminish its performance.
The above listed properties
make NN's interesting for {\it distributed} 
quantum information (QI), 
where quantum bits do not correspond to the internal states (spins) of individual ions, 
but 
to patterns of the internal states of the whole chain (all-up, all-down, half-up-half-down, etc.). 
These patterns {\it echo} the lowest energy vibrational modes of the system. The sign of the displacement of each ion with respect to its equilibrium position in a given mode fixes the up/down state of the spin in the associated spin pattern.
Some approaches to exploit the potential of NN models 
for QI processing have been 
discussed 
\cite{nn}, also with respect to entanglement generation \cite{sen}.  Here we propose, 
for the first time, a feasible implementation of NN, 
and the realization of distributed QI 
using a chain of trapped ions. 

We first remind
the readers the main features of the Hopfield model
\cite{hopfield}, and discuss its similarities with the effective Hamiltonian derived in
Ref. \cite{Wunderlich02,porras}, that suggest the possibility of using a chain of trapped ions as a NN. We find the ion-chain
storage capacity and its robustness the most appealing features of NN for distributed QI.
Thus, the question of ergodicity and, therefore, 
the ability of the system to act as an associative memory is not
considered here. 
We show that the storage capacity, which is determined by the phonon 
spectrum of the system, can be controlled by modifying the
shape of the external axial trapping potential and/or by applying
longitudinal magnetic fields. Although 
this is a
classical property of the network, spin ion systems also permit 
to study quantum NN by applying a transverse magnetic field or an
optical field that effectively simulates it. 
Here, we exploit the storage capacity of the system to perform distributed 
QI i.e. single and two-qubit gates by applying
appropriate external axial and transverse magnetic fields. 
Transverse magnetic fields should also permit tunneling
processes between stored patterns and to realize, for example,
quantum stimulated annealing \cite{parisi,ags}. 
This study is beyond the scope
of this Letter and will be treated elsewhere.

Following the models of Hopfield \cite{hopfield} and Little
\cite{little}, a neuron can be viewed as an Ising spin with two
possible states: ``up'' $(S=+1)$ and ``down'' $(S=-1)$ depending on whether the neuron has or has not fired an
electromagnetic signal, in a given interval of time
\cite{Amit}. The state of the network of $N$ neurons at a
certain time is defined by the instantaneous configuration of all
the spins $\lbrace S_i\rbrace$ at this time. The dynamic
evolution of these states is determined by the symmetric interactions among
neurons, $J_{ij}=J_{ji}$. 
Also, full connectivity is assumed,
i.e., every neuron can receive an input from any other one, and send 
an output to it. The Hamiltonian 
reads:
\begin{equation}
H=-\frac{1}{2}\sum_{i,j}^{N}J_{ij}S_i S_j + h\sum_{i}^{N}S_i,
\label{ham_neural}
\end{equation}
where $h$ corresponds to an external magnetic field.
The interactions are determined by
the patterns or configurations of spins to be stored in the network.
These patterns will be learned if the system is able to accommodate
them as attractors, implying that a large set of initial 
configurations of the network will be driven dynamically 
to those patterns. 
A possible choice of the interactions is
\begin{equation}
J_{ij}=\frac{1}{N}\sum _{\mu=1}^p \xi_i^{\mu}\xi_j^{\mu},
\label{int_neural}
\end{equation}
with $i\not= j$. The $p$ sets of $\lbrace \xi_{i}^{\mu}\rbrace = \pm 1$ are the patterns to be stored. 
The network will have the capacity of storage and retrieval of information, if the dynamical stable configurations
(local minima) reached by the system $\lbrace S_{i} \rbrace$ are correlated with the learned ones $\lbrace \xi_{i}^{\mu}\rbrace$. 
Although the interactions have been constructed to guarantee that 
certain specified patterns are fixed points of the dynamics, the non-linearity of the dynamical process may induce 
additional attractors, the so-called spurious states.

Recently it has been shown that the Hamiltonian describing a linear
chain of harmonically trapped ions exposed to a magnetic field
gradient \cite{Wunderlich02} or interacting with convenient laser fields \cite{porras} can be transformed into an effective spin-spin hamiltonian with LR interactions ($J^{\alpha}_{ij}$), mediated by the collective motion of
the ions:
\begin{equation}
H=-\frac{1}{2}\sum_{\alpha,i,j}J^{\alpha}_{ij}\sigma^{\alpha}_i\sigma^{\alpha}_j+\sum_{\alpha,i}B^{\alpha}_i\sigma^{\alpha}_i,
\label{hamiltonian}
\end{equation}
\begin{equation}
J^{\alpha}_{ij}=\frac{(F^{\alpha})^2}{m}\sum_n
\frac{M^{\alpha}_{i,n}M^{\alpha}_{j,n}}{\omega^2_{\alpha,n}},
\label{int}
\end{equation}
with $\alpha= x, y, z$, $(i, j)$ label the ions, $\sigma$ are the Pauli matrices, 
$F^{\alpha}$ the force in the $\alpha$ direction experimented by the ions, 
$m$ the ion mass and $\omega_{\alpha,n}$ the angular frequency of the vibrational mode $n$.
$M^{\alpha}_{i,n}$ are the
unitary matrices that diagonalize the vibrational Hamiltonian:
$M^{\alpha}_{i,n}{\kappa}^{\alpha}_{i,j}
M^{\alpha}_{j,m}=\omega^2_{\alpha,n} \delta_{nm}\,,$
where ${\kappa}^{\alpha}_{i,j}$ are the elastic constants of the
chain \cite{goldstein}. The coefficient $M^{\alpha}_{i,n}$ gives the
scaled amplitude of the local oscillations of ion $i$ around its
equilibrium position, when the collective vibrational mode $n$ is
excited. Thus, the eigenvectors of $M$ describe each ion's
contribution to a given vibrational mode, while the eigenvalues
provide the frequencies, $\omega_{\alpha,n}$ of the
collective modes.

The external trapping frequencies are chosen such that the laser
cooled ions crystallize in a linear chain (i.e.
$\omega_{x,1}=\omega_{y,1}\gg \omega_{z,1}$) and the external 
forces act on the $z-$axis (i.e.
$F^{x}=F^{y}=0$), so that the index $\alpha$ is
dropped. Henceforth, we consider zero magnetic fields
$B_{i}=0$ \cite{noteta}. 
If we substitute the Pauli matrices in Eq.(\ref{hamiltonian}) 
by Ising spins $S=\pm 1$ (where the effective spin corresponds to the 
internal state of the ion), we recover Eq.(\ref{ham_neural}) and the possibility to
implement a classical NN with this system arises. 
Nevertheless, there are some differences
between both models. First, in the Hopfield model, the interactions
(Eq.(\ref{int_neural})) are determined by the patterns to be
stored $\lbrace \xi_{i}^{\mu}\rbrace=\pm 1$, while in the trapped ion
chain, the interactions are fixed by the
collective modes of the system, i.e., the coefficients $M_{i,n}$ 
that do not necessarily equal $\pm 1$. Second,
in Eq.(\ref {int_neural}), $p$ corresponds to the number of
patterns to be stored, which in the limit of large $N$ (number of spins),
is bounded from above by $p=0.14N$  \cite{Amit}. In 
Eq.(\ref{int}), the sum extends over the total number of
vibrational modes which is larger than the total number of
stored patterns (spin configurations that the system is able to recover).
And finally, in the Hopfield model all the
patterns have the same weight while in the ion chain each
vibrational mode is weighted by $1/ \omega^{2}_{n}$ (Eq.(\ref{int})). 
To reproduce as closely as possible a NN behavior, 
the most relevant requirement is the degeneracy 
of the vibrational modes. 
Moreover, the corresponding patterns must have large basins
of attraction, i.e. they should correspond to sufficiently
different spin configurations, so that each one is
dynamically recovered, even if several spins are randomly flipped.

To check the feasibility of implementing a NN model in these ion spin
systems,  
we first find the phonon spectrum using a standard diagonalization procedure, 
and impose the learning rule i.e., we
calculate the spin interactions $J_{ij}$, mediated by the collective modes of the ions (Eq.(\ref{int})). Then, we map a given vibrational mode into a spin 
configuration (initial spin configuration), evaluate its energy (Eq.(\ref{hamiltonian})) and check its dynamical stability under spin flips using
a standard Metropolis algorithm in a classical Monte Carlo code. 
This stability is essential for adiabatic QI processing. 
Explicitly, from the initial configuration we randomly flip $r$ spins, and let the
system evolve towards equilibrium assuming a noiseless scenario.
If the system recovers the initial configuration, it is stable
under the flip of $r$ spins. 
We define the initial overlap as $m_i=(N-r)/N$. After
dynamical evolution, the final overlap is given by $m_j=(N-s)/N$
where $s$ is the number of spins that differ from the initial
configuration. We repeat this process over $M$ initial configurations
each with  $r$ random spin flips.  We evaluate statistically the final overlap
with the initial configuration as: $m_f=(\sum_{j=1}^N m_j n_j)/M$, being $n_j$ the number of times that the system reaches 
the configuration with $m_j$. 
The value of $m_i$ for which significant decrease of $m_f$ occurs, is a good measure of the size of the pattern's basin of attraction.
For the harmonic trapping potential, the two lowest vibrational modes are the
center of mass (CM) (all spins parallel, with $\omega_1$) 
and the breathing (B) mode (half up, half down, with $\omega_2=\sqrt{3} \omega_1$) \cite{james}.
We find that the pattern associated with the CM mode is stable up to the flip of
half of the spins, while the one associated with the B mode is already unstable under
a single spin flip. Thus, only the spin configuration associated with the CM mode can be recovered (i.e., stored).
\begin{figure}
\includegraphics[width=0.85\linewidth,height=6.70cm]{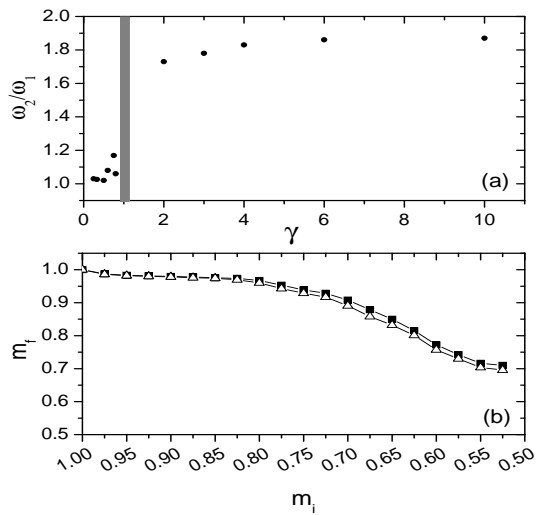}
\caption{(a) Ratio between the frequencies of the second and first vibrational modes as a function of the exponent of the trapping potential for 20 Ca$^+$ ions. (b) Final overlap averaged over 500 initial configurations 
as a function of the initial overlap, for the patterns associated with the two lowest vibrational modes of 40 ions in a potential $V=\rho|x|^{0.5}$ with $\rho=6.6\times 10^{-20}J/m^{1/2}$. The black squares (triangles)
correspond to the first (second) pattern.}
\label{fig1}
\end{figure}
To increase the storage
capacity of the network, 
we consider here $V(x)=\rho \mid x \mid^{\gamma}$, achievable using additional control 
by dc electrodes (\cite{christoph}(b), see also \cite{magnetic}). 
We calculate the ratio
$\omega_2/\omega_1$, as a function of
$\gamma$, for Ca$^+$ ions. For $N\geq 20$, this ratio depends 
neither on the number of ions, nor
on the value of $\rho$. For $\gamma \geq 1$,
the ratio is $\approx \sqrt{3}$, and as in the harmonic case, severe limitations on the storage capacity appear (see Fig.\ref{fig1}(a)). 
However, for $0.25< \gamma < 0.8$, the two lowest modes become nearly degenerate. 
The storage capacity for a system of 40 Ca$^+$ ions trapped in a potential with $\gamma=0.5$ is displayed in Fig. \ref{fig1}(b), where the final overlap is depicted as a
function of the initial overlap for the patterns associated with the two lowest vibrational modes. 
$m_f$ is close to 1 up to 8 spin flips, meaning that the system is able to
recover four patterns (the two associated with the two lowest modes plus the two corresponding to a global  spin flip).
The system sometimes reaches a slightly deformed configuration, 
which differs only in 1 spin flip from the original one 
(spurious states), making $m_f$ slightly smaller than 1. 
Specifically, the probability of recovering the two modes is above 
 $98\%$, up to 3 initial 
random spin flips, and above $97\%$ up to 8 ($m_i=0.8$).

Having shown that our system can be robust for classical information storage,
%
we explore now 
its capability for \emph{distributed}
quantum computing in an error resistant way (cf. \cite{web}), i.e. robust with respect to the partial damage of the system. To this aim, we consider a system 
of 8 spin-1/2 particles (as in \cite{ashtam}) in a trapping potential
with $\gamma\simeq 0.5$. The vibrational spectrum is, except for the lowest
two modes, highly non-degenerate, with rapidly increasing eigenvalues. Thus, we consider only the contributions of the spin configurations associated with the three lowest motional modes which, 
up to a gauge transformation, correspond to: all-up, half-up-half-down, 2-up-4-down-2-up. We encode the information in the first two, and consider the third one as noise. The phonon mode amplitudes are approximated as $M_{i,m}=\xi^m_i=\pm 1$ (exact for periodic 
boundary conditions). Additionally, time dependent ``magnetic'' fields 
in the $z$ ($B_1$, $B_2$) and $x$ ($A$) directions are applied leading to the following expression
for a Quantum Neural Network (QNN) Hamiltonian:
\begin{eqnarray}
&&H_{QNN}(t) = - \lambda \Big[r_1 (S^{z}_1+S^{z}_2+S^{z}_3+S^{z}_4)^2  + \nonumber \\
&&r_2(S^{z}_1+S^{z}_2-S^{z}_3 - S^{z}_4)^2 
+r_3(S^{z}_1- S^{z}_2-S^{z}_3+S^{z}_4)^2   \nonumber \\
&&+A(S^{x}_1+S^{x}_2+S^{x}_3+S^{x}_4) \nonumber \\
&&+B_1(S^{z}_1+S^{z}_2)+B_2(S^{z}_3+S^{z}_4)\Big], 
\end{eqnarray}
where \(S^{\alpha}_i= \sigma^\alpha_{2i-1} + \sigma^\alpha_{2i}\), \(i=1,\ldots,4\), and $r_1\approx r_2\gg r_3>0$. With \(\lambda\) in energy units, all the other parameters in \(H_{QNN}\) are dimensionless. $B_1$, $B_2$, and $A$ are chosen 
initially ($t=t_0$) such that the ground \(|G(t_0)\rangle\), 
and the first three excited states \(|E_j(t_0)\rangle\) correspond to:
$|\uparrow\uparrow\uparrow\uparrow\uparrow\uparrow\uparrow\uparrow\rangle_z$, 
$|\downarrow\downarrow\downarrow\downarrow\downarrow\downarrow\downarrow\downarrow\rangle_z$,
$|\uparrow\uparrow\uparrow\uparrow\downarrow\downarrow\downarrow\downarrow\rangle_z$, and $|\downarrow\downarrow\downarrow\downarrow\uparrow\uparrow\uparrow\uparrow\rangle_z$. 
\begin{figure}
\includegraphics[width=1.0\linewidth,height=4cm]{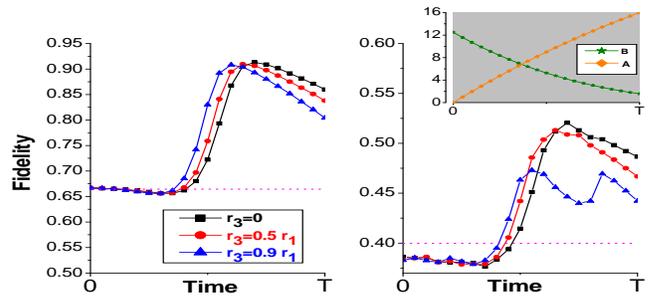}
\caption{(Color online)
Fidelities of the \(\cal H\) gate (left) and the Bell gate (right),
 with  respect to time (\(r_2 = 0.95 r_1\)). The classical fidelity bounds are \(2/3\) and \(2/5\), respectively (horizontal dashed lines). 
The inset in gray shows  the fields \(A \lambda\) and \(B_1 \lambda = 10^{-5} B \lambda\) and \(B_2 \lambda = 10^{-6}  B \lambda\) 
with respect to time. For adiabaticity, the chosen fields  require \(T \gg 7 \times 10^6 \hbar/\lambda\).
The fidelities are largely independent of the actual dynamics of the fields.}
\label{fig_kanak}
\end{figure}
  
To demonstrate universality we focus first on a single qubit operation. 
Identifying $|0\rangle=|G(t_0)\rangle$, $|1\rangle=|E_1(t_0)\rangle$, 
we consider a single distributed-qubit operation, with a generic qubit \(a_0|G(t_0)\rangle + a_1|E_1(t_0)\rangle\), evolving
adiabatically under the changes of the magnetic fields,
to a final state \(a_0|G(t=T)\rangle + a_1|E_1(t=T)\rangle\). We choose the
 time dependence of the fields so that the final state is approximately
the \({\cal H}\)-rotated state of the input one, where
\({\cal H}\) acts on the logical states as \(|0\rangle \to |+\rangle\),
 \(|1\rangle \to -|-\rangle\), where \(|\pm\rangle = (|0\rangle \pm |1\rangle)/\sqrt{2}\). 
This one-qubit gate \({\cal H}\) is reminiscent of the 
 Hadamard gate that takes \(|0\rangle \to |+\rangle\),
 \(|1\rangle \to |-\rangle\). The $\cal H$ operation is achieved by 
changing the positive initial values of \(B_1\) and 
\(B_2\) adiabatically to zero, and simultaneously increasing 
the zero initial $A$ to a positive value 
much larger than the initial \(B_1,B_2\) (inset Fig. \ref{fig_kanak}).
Since we deal with superpositions of energy eigenstates, 
we consider in the adiabatic transport
the dynamical, as well as the Berry phases \cite{Pancharatnam}. 
The fidelity of the $\cal H$ gate is shown 
in Fig. \ref{fig_kanak}(left) as a function of time for different noise
ratio \(r_3/r_1\). 
Note that artificially increasing the ratio \(r_3/r_1\) imitates inaccuracies in the trapping potential,
disturbance in the motion of the ions, as well as noise in the spin (as the phonons are the carriers of interaction between the spins).
The fidelity is quite insensitive to high noise levels, 
and is larger than the classical (measure and prepare) bound of \(2/3\). 

Let us move now to the two-qubit gates, and treat the 4 left spins as one qubit, and 4 right ones as another, so that:
$|00\rangle=|\uparrow \uparrow \uparrow \uparrow \uparrow\uparrow \uparrow \uparrow\rangle_z$, 
$|01\rangle=|\uparrow \uparrow \uparrow \uparrow\downarrow \downarrow \downarrow \downarrow\rangle_z$, etc. 
(We have checked that the \({\cal H}\) gate fidelity is robust in this encoding also.)
 We demonstrate here, a way to implement an entangling universal gate \cite{hazar} acting as 
\(|00\rangle \to (|00\rangle + |11\rangle)/\sqrt{2}\), \(|11\rangle \to (-|00\rangle + |11\rangle)/\sqrt{2}\),
\(|01\rangle \to (|01\rangle + |10\rangle)/\sqrt{2}\), \(|10\rangle \to (-|01\rangle + |10\rangle)/\sqrt{2}\). 
We denote this gate as Bell gate. We now encode an arbitrary two-qubit state \(\sum_{i,j=0,1}a_{ij}|ij\rangle\), into 
\(a_{00}|G(t_0)\rangle + a_{11}|E_1(t_0)\rangle + a_{01}|E_2(t_0)\rangle + a_{10}|E_3(t_0)\rangle\). The same variation of 
the magnetic fields as in the $\cal H$ gate leads now to the Bell-gate rotated state (Fig. \ref{fig_kanak} (right)), with fidelity 
that is noise insensitive and surpasses the classical bound  of  \(2/5\). 
Note that in addition to being resistant against noise induced by increasing \(r_3/r_1\), the fidelities are also robust against 
spin-flip errors, as we have encoded the qubit(s) in the two (four) lowest energy levels, for the \({\cal H}\) (Bell) gate, 
which we have already shown to be metastable against spin-flips. 
The time scales for which both \({\cal H}\) and 
Bell gate fidelities reach maximum values are long enough to ensure 
robust implementation and also robustness against errors in time of observation.

Summarizing, we have shown that spin-ion systems can be used to implement
NN models. We have calculated their storage capacity and 
robustness against spin flips as well as 
their dependence on the trapping potential. 
Identifying the qubits with configurations of spins that {\it echo} the lowest vibrational modes of the system, we have shown that the system 
can perform error resistant universal distributed QI 
processing. We have demonstrated that by applying adiabatically-varying time dependent ``magnetic'' fields, the system 
realizes single and two distributed-qubit operations in a robust way \cite{fahri}. 
The scalability issue is like other proposals
and experiments in ion-trap quantum computing \cite{barey-barey-phirey-asey}, 
and may potentially be overcome by connecting mesoscopic clusters of 
trapped ions for instance, by ``flying'' qubits.

We thank  I. Bloch, A. Bramon, H.-P. B\"uchler, J. Eschner, M. Mitchell, 
J. Wehr, P. Zoller for fruitful discussions, and
acknowledge  German DFG
SFB 407,
ESF PESC QUDEDIS, 
Spanish MEC (FIS2005-04627;FIS2005-01369, Consolider-Ingenio2010 CSD2006-00019), AvH Foundation, 
EU IP SCALA and QAP for support.


\begin{thebibliography}{99}

\bibitem{cirac} J.I. Cirac and  P. Zoller, Phys. Rev. Lett. \textbf{74}, 4091 (1995).

\bibitem{general} D. Bouwmeester, A. Ekert, and A. Zeilinger (Eds.), \emph{The Physics of Quantum Information} (Springer, Berlin, 2000).

\bibitem{experiments}  F. Schmidt-Kaler \emph{et al.}, Nature \textbf{422}, 408 (2003);
J. Chiaverini \emph{et al.}, Science \textbf{308}, 997 (2005); M. Riebe \emph{et al.},
 Nature \textbf{429}, 734 (2004); M. D. Barrett \emph{et al}, Nature \textbf{429}, 737 (2004). 



\bibitem{christoph} (a) F. Mintert and C. Wunderlich, Phys. Rev. Lett. \textbf{87}, 257904
(2001); Ch. Wunderlich and Ch. Balzer, Adv. At. Mol. Opt. Phys. {\bf
49}, 293 (2003); (b) D. McHugh and J. Twamley, Phys. Rev. A {\bf 71},
012315 (2005).
\bibitem{Wunderlich02} C.\ Wunderlich,  in {\em
Laser Physics at the Limit} (Springer, Heidelberg, 2002);
also available as 
quant-ph/0111158.



\bibitem{porras}D. Porras and J.I. Cirac, Phys. Rev. Lett. \textbf{92}, 207901 (2004).
\bibitem{Amit} D.J. Amit,  \textit{Modeling Brain Function} (Cambridge
 University Press, 
Cambridge, 1989).
\bibitem{parisi} M. M\'ezard, G. Parisi, and M.A. Virasoro, {\it Spin Glass Theory and Beyond} (World Scientific, Singapore, 1987).
\bibitem{hopfield} J.J. Hopfield, Proc. Natl. Acad. Sci. \textbf{81}, 3088 (1984).
\bibitem{nn} P. Gralewicz, quant-ph/0401127, and references therein.
\bibitem{sen} A. Sen(De) \emph{et al.}, Phys. Rev. A \textbf{74}, 062309(2006).
\bibitem{ags}  D.J. Amit, H. Gutfreund and  H. Sompolinsky, Phys. Rev. A \textbf{32}, 1007 (1985).
\bibitem{little} W.A. Little, Math. Biosci. \textbf{19}, 101 (1974).
\bibitem{goldstein} H. Goldstein, \emph{Classical Mechanics} (Addison-Wesley,  MA, 1980).

\bibitem{noteta} The effective magnetic field $B$ in Eq.
(\ref{hamiltonian}) includes a contribution from the applied lasers, 
which can be compensated by a constant magnetic field $B'$ \cite{porras}:
$B=B'+\frac{F^2}{m \omega_1^2}$.

\bibitem{james} D.F.V. James, Appl. Phys. B \textbf{66}, 181 (1998); G. Morigi 
and S. Fishman,
Phys. Rev. Lett. {\bf 93}, 170602 (2004).
\bibitem{magnetic} The storage capacity of the system can be also increased by engineering the
spatial dependence of a longitudinal magnetic field. 
\bibitem{web}http://www.perimetertinstitute.ca/personal/dgottesman/ and references therein.
\bibitem{ashtam} H. H\"affner \emph{et al.}, Nature {\bf 438}, 643 (2005).
\bibitem{Pancharatnam} A. Shapere and F. Wilczek, {\it Geometric Phases in Physics}, (World Scientific, Singapore, 1998).
\bibitem{hazar} D. Deutsch, A. Barenco and A. Ekert, Proc. Roy. Soc. London A {\bf 449}, 669 (1995). 
\bibitem{fahri} It has certain similarities, but also differences with adiabatic 
quantum computation. See e.g. E. Farhi and S. Gutmann, Phys. Rev. A {\bf 57},  2403 (1998).
\bibitem{barey-barey-phirey-asey} See e.g., S. Seidelin \emph{et al.}, quant-ph/0601173.
\end{thebibliography}
\end{document}